\documentclass[]{article}
\usepackage{proceed2e}
\usepackage{amsmath,bm}
\usepackage{amssymb}
\usepackage{graphicx}
\usepackage{subfigure}
\usepackage{natbib} 
\usepackage{pzccal}

\def \Btau   { {\mbox{\boldmath $\tau$}}}

\def \Bf { {\bf f} }
\def \Bg { {\bf g} }

\def \Bt { {\bf t} }
\def \Bu { {\bf u} }

\def \Bw { {\bf w} }
\def \Bx { {\bf x} }
\def \By { {\bf y} }
\def \Bz { {\bf z} }

\def \BB { {\bf B} }

\def \BG { {\bf G} }

\def \BK { {\bf K} }

\def \BM { {\bf M} }

\def \BP { {\bf P} }

\def \BS { {\bf S} }
\def \BT { {\bf T} }

\def \BW { {\bf W} }

\newcommand{\ycost}{\ensuremath{{\tilde{q}(\By_t)}}}
\newcommand{\xcost}{\ensuremath{{{q}(x_t)}}}
\newcommand{\ysys}{\ensuremath{\mathpzc{y}}}
\newcommand{\app}{\ensuremath{_\text{ap}}}
\newcommand{\klexact}{{\bf KL (exact) }}
\newcommand{\vkl}{{\bf VKL }}
\newcommand{\avkl}{{\bf AVKL }}

\title{Latent Kullback Leibler Control for Continuous-State Systems \\ using Probabilistic Graphical Models} 

\author{} 

\author{ {\bf Takamitsu Matsubara$^{\dagger}$}
\And 
{\bf Vicen\c{c} G\'omez$^{\S,*}$} \vspace{.1cm} \\ 
$^{\dagger}$ Graduate School of Information Science. Nara Institute of Science and Technology (NAIST). Nara, Japan\\
$^{\S}$ Donders Institute for Brain, Cognition and Behaviour. Radboud University Nijmegen, the Netherlands\\
$^{*}$ Department of Information and Communication Technologies. Universitat Pompeu Fabra. Barcelona, Spain\And 
{\bf Hilbert J. Kappen$^{\S}$}\\
} 
 
\begin{document} 
\bibliographystyle{apalike}


\maketitle 

\begin{abstract} 
Kullback Leibler (KL) control problems allow for efficient computation of optimal control by solving a principal eigenvector problem.
However, direct applicability of such framework to continuous state-action systems is limited. In this paper, we propose to embed a KL control problem in a probabilistic graphical model where observed variables correspond to the continuous (possibly high-dimensional) state of the system and latent variables correspond to a discrete (low-dimensional) representation of the state amenable for KL control computation.
We present two examples of this approach. The first one uses standard hidden Markov models (HMMs) and computes exact optimal control, but is only applicable to low-dimensional systems. The second one uses factorial HMMs, it is scalable to higher dimensional problems, but control computation is approximate. We illustrate both examples in several robot motor control tasks. 
\end{abstract} 

%
%

\section{INTRODUCTION}
\label{sec1}
Recent research in stochastic optimal control theory has identified a class
of problems known as Kullback-Leibler (KL) control problems \citep{KappenML2012} or linearly solvable Markov decision problems (LSMDPs) \citep{TodorovNIPS2007}.
For these (discrete) problems, the set of actions and the cost function are restricted in a way that makes the Bellman equation linear and thus more efficiently solvable, for instance,
by solving the principal eigenvector of a certain linear operator \citep{TodorovPNAS2009}.  

However, direct applicability of this framework to continuous state-action systems, such as robot motor control, is limited. The main problem is the curse of dimensionality, which appears because discretization quickly leads to a combinatorial explosion. 
This problem has been addressed using function approximation methods in \citep{TodorovADPRL2009}. Instead of directly solving a discrete-state LSMDP, these methods approximate the so-called desirability function, which is defined in the continuous-state space. \citet{KinjoFN2013} combined this function approximation scheme with system identification on a real robot navigation task. However, approaches based on the continuous-state formulation of KL control problems have several limitations: they require to solve a quadratic programming problem, a more computationally demanding problem than computing the principal eigenvector. Also, there is no guarantee of convergence to a positive solution.  Alternative formulations that address these limitations have been recently proposed \citep{ZhongIFAC2011,Zhong2011Moving}.  \citet{ZhongIFAC2011} used a soft aggregation method to solve KL-control problems in an aggregated space.  Both approaches, however, require the model of system dynamics, which is often not available in real-world applications \citep{KinjoFN2013}. 

In this paper, we propose to embed a KL-control problem in a probabilistic graphical model with mixed continuous and discrete variables. The continuous variables correspond to the (possibly high-dimensional) state of the system and the discrete variables correspond to a latent (low-dimensional) representation of the state which is amenable for KL control computation. The model parameters are first learned using data from the real system running with exploring controls. The control input to the real system is then computed as a filtering step combined with the solution of the KL-control problem in the latent space.

We present two examples of this approach: the first one uses a standard hidden Markov model (HMM) in which inference can be computed exactly, but is only applicable to low-dimensional continuous systems. The second one uses factorial HMMs (FHMMs) and is applicable to higher dimensional problems, although optimal control can only be approximated. We illustrate both examples in several robot motor control tasks. In particular, 
we experimentally demonstrate that the second example with FHMMs is scalable to high-dimensional problems (e.g., 25 dimensional problem) that may not be solvable by other approaches.

\section{KULLBACK LEIBLER CONTROL PROBLEMS}
\label{sec:klcontrol}
We briefly summarize the class of KL control problems introduced by \citet{TodorovNIPS2007} in
the infinite-horizon average-cost formulation \citep[see also][]{TodorovPNAS2009}.

Let $\mathcal{X}=\{1,\hdots,N\}$ be a finite set of states and $\mathcal{U}(x)$ be
a set of admissible control actions at state $x \in \mathcal{X}$.
Consider the transition probability $p(x'|x)$ that describes the system dynamics in the absence of 
control. Such {\it uncontrolled dynamics} assigns zero probability for physically forbidden state transitions. Denote the transition probability given action $u \in \mathcal{U}(x)$ as $p(x'|x,u)$
and the immediate cost for being in state $x$ and taking action $u$ as $\ell(x,u)\geq 0$.

For infinite-horizon problems, the objective is to find a control law
$u=\pi(x)$ that minimizes the average cost
\begin{align}\label{eq:avgcost}
\lim_{n\rightarrow\infty} \frac{1}{n} \mathbb{E}\left[\sum_{t=0}^{n-1} \ell(x_t,\pi(x_t))\right]
= \sum_x \Pi(x)\ell(x,\pi(x))
\end{align}
where $n$ is the number of time-steps and $\Pi(x)=\lim_{t\rightarrow\infty}p(x_t=x|x_0,\pi)$ is the stationary distribution of states under control law $\pi$, which we assume exists and is independent of $x_0$, i.e., $p(x_t=x|x_0,\pi)$ is assumed ergodic. 

The following Bellman equation defined for the (differential) cost-to-go function $v(x)$ minimizes Eq.~\eqref{eq:avgcost}
\begin{align}\label{Eq:Belman}
c + v(x) = \min_{u \in \mathcal{U}(x)} \left\{ \ell(x,u) + \mathbb{E}_{x'\sim p(\cdot|x,u)}[v(x')]\right\},
\end{align}
where $c$ is the average cost that does not depend on the starting state.

Minimizing Eq.~\eqref{Eq:Belman} is in general hard, but in some cases it can be done efficiently.
KL control problems are a class of problems for which Eq.~\eqref{Eq:Belman} becomes linear under the following assumptions:

\noindent \emph{(i)} the controls directly specify state transition probabilities, i.e. $p(x'|x,u) = u(x'|x)$. The action vector $u(\cdot|x)$ is a probability distribution over next states given the current state $x$.

\noindent \emph{(ii)} the immediate cost function has the following form
\begin{align*}
\ell(x,u) = \alpha q(x) + {\rm KL} \left(u(\cdot|x) \parallel p(\cdot|x)\right),
\end{align*}
where $q(x) \geq 0$ is an arbitrary state-dependent cost and ${\rm KL}$ is the Kullback Leibler divergence between the controlled and the uncontrolled dynamics, reflecting how much the control changes the normal behavior of the system.
Parameter $\alpha$ allows to balance the two cost terms. 

Define the exponentiated cost-to-go (desirability) function  $z(x)=\exp(-v(x))$ and the linear operator
\begin{align*}
\mathcal{G}[z](x) & = \sum_{x'}p(x'|x)z(x')= \mathbb{E}_{x'\sim p(\cdot|x,u)}[z(x')].
\end{align*}
The resulting minimization takes the form
\begin{align*}
\min_{u\in\mathcal{U}(x)}\hspace{-.1cm} \left\{ \alpha q(x)\hspace{-.1cm} + {\rm KL} \left(u(\cdot|x) \left\|\frac{p(\cdot|x)z(\cdot)}{\mathcal{G}[z](x)}\right.\right)\hspace{-.1cm}-\log\mathcal{G}[z](x)\hspace{-.05cm}\right\}.
\end{align*}
At the global minimum, the Bellman equation becomes
\begin{align*}
\exp(-c)z(x) &= \exp(-\alpha q(x))\mathcal{G}[z](x) 
\end{align*}
or in matrix form
\begin{align}\label{eq:linear}
\lambda \Bz & = \BG \BP \Bz
\end{align}
where $\BG$ is a $N\times N$ diagonal matrix with elements $\exp(- \alpha q(x))$ and $\lambda = \exp(-c)$. 
From Eq.~\eqref{eq:linear}, it follows that $\Bz$ is any eigenvector of the matrix 
$\BG \BP$ with eigenvalue $\lambda$. The optimal average cost becomes 
$c = - \ln \lambda$. Thus, the minimal solution is given by the principal 
eigenvector of $\BG \BP$: the eigenvector $\Bz^*$ with largest eigenvalue, which can be efficiently computed using the power iteration method
\citep{TodorovNIPS2007}. The optimal control is given by 
\begin{align}\label{eq:optu}
u^*(x'|x) &= \frac{p(x'|x) \Bz^*(x')}{\mathcal{G}[\Bz^*](x)}. 
\end{align}
%

\section{LATENT KULLBACK LEIBLER CONTROL}
\label{sec3}
The previously described framework is not directly applicable for continuous systems.
For such cases, we propose to learn a discrete hidden representation and dynamics amenable for efficient computation from the observed continuous variables. Our approach can be summarized in the following three steps:
\begin{enumerate}
\item Learn a probabilistic graphical model from data samples obtained for the real system
\item Solve the KL control problem in the latent space of the probabilistic graphical model
\item Compute control in the observed space
\end{enumerate}
This general method is directly applicable to arbitrary continuous state-action systems, 
while in this paper we focus on the following deterministic control-affine systems that typically describe discrete-time robot dynamics: 
\begin{align}\label{eq:dynamics}
\ysys_{t+1} & = \ysys_t  + \Delta t \left( \Bf(\ysys_t) + \BB(\ysys_t) \Btau_t\right),
\end{align}
where $\ysys_t \in \mathbb{R}^{D}$ is the state variable of the
system, $\Btau_t \in \mathbb{R}^{d}$ is the control input, $\Bf(\ysys_t) \in
\mathbb{R}^{D}$ is the uncontrolled dynamics, $\BB(\ysys_t) \in \mathbb{R}^{D \times
d}$ is the control matrix and $\Delta t$ is the discrete-time step-size. 

Two particular realizations of this general approach are described in the next section. 
The first one uses standard HMMs, which are the most natural way to model sequences of observations. 
However, it is only applicable to systems in which the relevant region of the state-space is small, such as low-dimensional systems, or largely constrained high-dimensional systems.
The second one uses factorial HMMs, which assume factorized uncontrolled dynamics and can scale up to higher dimensional problems. 


\section{EXACT CONTROL COMPUTATION USING HIDDEN MARKOV MODELS}
\label{sec:hmm}
In this section, we describe an example of latent KL control based on standard hidden Markov models.


\subsection{LEARNING HMMS FOR KL CONTROL}
\label{subsec}
Consider the hidden Markov model with hidden states $x_t \in \{1,\hdots,N\}$, stochastic state transition matrix $\BP$ with entries $P_{ij}=p(x_{t+1}=j|x_t=i)$ and Gaussian observation model $p(\By_t|x_t=k) = \mathcal{N}(\mu_k,\Sigma_k)$.

We generate sample trajectories $\mathcal{D} = \{\By_t, \hdots, \By_T\}$ from the real system driven solely by exploration noise (uncontrolled dynamics) and use them to learn the parameters $\boldsymbol{\theta}_{\text{HMM}} = \{\BP,\mu_{1:N}, \Sigma_{1:N}\}$. After learning, the matrix $\BP$ encodes a coarse description of the observed dynamics in a latent space and the Gaussian means and variances capture the relevant regions in this space.
More precisely, considering the system of Eq.~\eqref{eq:dynamics}, we set exploration noise as $\Btau_t = \epsilon_t$ for $t=1\hdots T$, where $\epsilon_t \in \mathbb{R}^{d} \sim \mathcal{N}(0,\Sigma_{\epsilon})$. 
The choice of such a zero-mean Gaussian distribution is motivated by the relationship between the KL action cost and the input-norm cost: in the continuous setting the KL cost reduces to a quadratic energy cost \citep{TodorovPNAS2009, KappenML2012}, which coincides with a commonly used input-norm cost for energy-efficient or smooth motor control behavior \citep{MitrovicICRA2009}. 
The covariance matrix $\Sigma_{\epsilon}$ is a free
parameter. For low exploration noise, one would expect the learned model to be
a poor approximation since only a small fraction of the state space is visited.
Conversely, large noise values would result in too flexible models with
unrealistic state transitions. The correct noise value is therefore a trade-off
between these two scenarios.

Given $\mathcal{D}$, the parameters $\boldsymbol{\theta}_{\text{HMM}}$ can be learned, for instance, using the standard Expectation-Maximization (EM) algorithm (Baum-Welch algorithm).

%
%

\subsection{CONTROL COMPUTATION IN LATENT SPACE}
\label{sec:costhmm}
To define a KL control problem in the latent space, we first need a state-dependent cost function expressed in terms of the latent variable $x$.
Let $\ycost$ and $\xcost$ be the cost functions in observation and latent
spaces, respectively.  We define $\xcost$ given $\ycost$ using $\exp(-
\xcost) = \int_{\By_t} \exp(-\alpha \ycost) p(\By_t|x_t) d\By_t$.  Furthermore,
if $\ycost$ is given in quadratic form $\ycost = (\By_t - \mu_{q})^T
\Sigma_{q}^{-1} (\By_t - \mu_{q})=
||\By_t - \mu_{q}||^2_{\Sigma_{q}^{-1}}$ 
and the observation model is 
Gaussian $p(\By_t|x_t) = \mathcal{N}(\mu_x, \Sigma_x)$, we can obtain $\xcost$ analytically:
\begin{align*}
\xcost & = - \ln \left \{\int_{\By_t} \exp\left( - \alpha \ycost \right) p(\By_t|x_t) d\By_t \right \}\\
& = - \ln \left \{\frac{|\BS|^{1/2}}{|\Sigma_x|^{1/2}} \exp\left[ - \frac{1}{2} ||\mu_q - \mu_x||^2_{\BM^{-1}} \right] \right \} \nonumber 
\end{align*}
where, 
$\BS = (\alpha \Sigma_q^{-1} + \Sigma_x^{-1})^{-1}$ and $\BM = \alpha^{-1} \Sigma_q + \Sigma_x$.

The (latent) KL control problem can now be formulated using state cost $q(x_t)$ and uncontrolled dynamics
$\BP$ as in Eq.~\eqref{eq:linear}. The optimal state transition $u^*(x_{t+1}|x_t)$ under controlled dynamics is given by Eq.~\eqref{eq:optu}.
%


\subsection{CONTROL COMPUTATION IN OBSERVED SPACE}
\label{sec:optcont}
We are now ready to describe how to use latent KL control in the real system. Given an observation sequence $\By_{1:t}$ until time $t$, we can compute predictive distributions of the next observation $\By_{t+1}$ under both the uncontrolled dynamics $p(x_{t+1}|x_{t})$ and the optimally controlled dynamics $u^*(x_{t+1}|x_t)$ in the latent space as: 
\begin{align*}
p(\By_{t+1}|\By_{1:t}) & = \sum_{x_{t:t+1}} p(\By_{t+1}|x_{t+1})  p(x_{t+1}|x_{t}) u(x_t|\By_{1:t})\\
u(\By_{t+1}|\By_{1:t}) & = \sum_{x_{t:t+1}} p(\By_{t+1}|x_{t+1})  u^*(x_{t+1}|x_{t}) u(x_t|\By_{1:t})
\end{align*}
where $u(x_t|\By_{1:t})$ denotes the filtered state at time $t$ following the controlled process that evolves
according to $u^*(x'|x)$. Since we keep the previous filtered estimate 
$u(x_{t-1}|\By_{1:t-1})$, this computation is simply as 
\begin{align*}
u(x_t|\By_{1:t}) & = \frac{p(\By_{t}|x_t) \sum_{x_{t-1}} u^*(x_{t}|x_{t-1}) u(x_{t-1}|\By_{1:t-1})}  {u(\By_{t}|\By_{1:t-1}) }. 
\end{align*}

We finally compute the control input command to the system such that the ``difference'' between the uncontrolled and optimal behaviors is reduced
\begin{align}\label{eq:pd}
\Btau_{t} = \BK (\bar{\By}_{t+1|1:t}^{u} - \bar{\By}^{p}_{t+1|1:t}),
\end{align}
where $\bar{\By}_{t+1|1:t}^{u}$ and $\bar{\By}_{t+1|1:t}^p$ are 
the expectations of $\By$ over $u(\By_{t+1}|\By_{1:t})$ and 
$p(\By_{t+1}|\By_{1:t})$ respectively and $\BK$ is a gain matrix to be tuned. 
The gain $\BK$ can be optimally computed if the model of system dynamics is available \citep{TodorovADPRL2009}, 
however, in this paper we focus on the model-free scenario and leave it as a free parameter.

\section{APPROXIMATE CONTROL USING FACTORIAL HIDDEN MARKOV MODELS}
\label{sec4}
For high-dimensional problems that require to cover large regions of the state space, the previous approach becomes infeasible, since the cardinality required for the latent variable grows exponentially. In this section, we consider an alternative model with a multi-dimensional latent variable and constrained state transitions. We consider each dimension independent from the rest in the absence of control. These assumptions are naturally expressed using factorial HMMs. The advantage is that we can capture complex latent dynamics more efficiently. The price to pay is that exact optimal control computation in the latent space is no longer feasible and different approximation schemes have to be used. We describe this approach in the following sections.

\subsection{FACTORIAL HIDDEN MARKOV MODELS}
\label{subsec41}
FHMM is a special type of HMM to model sequences of observations originated from multiple latent dynamical processes that interact to generate a single output \citep{GhahramaniFHMM1997,MurphyBook2012}.
The state is represented by a collection of variables $\Bx_{t} = \{x_t^{(1)},\hdots,x_{t}^{(m)},\hdots,x_{t}^{(M)}\}$ each of them having $K$ possible values. The latent state $\Bx_t$ is thus a $M$-dimensional variable with $K^M$ possible values.

We will use a $1$-of-$K$ encoding, such that each state component $\Bx_{t}^{(m)}$ will be denoted using a $K \times 1$ vector, where each of the $K$ discrete values corresponds to a $1$ in one position and $0$ elsewhere.

The assumption is that the transition model factorizes among the individual components
\begin{align}
p(\Bx_{t}|\Bx_{t-1}) = \prod_{m=1}^{M} p^{(m)}(\Bx^{(m)}_{t}|\Bx^{(m)}_{t-1}),
\end{align}
where $p^{(m)}(\Bx^{(m)}_{t}|\Bx^{(m)}_{t-1})$ is the state transition matrix $\BP^{(m)}$ for the $m$-th chain.
We assume the Gaussian observation model, which is defined as
\begin{align}\label{eq:obs-FHMM}
p(\By_t|\Bx_t) & = \mathcal{N} \left( \sum_{m=1}^M \BW^{(m)} \Bx_{t}^{(m)}, \Sigma  \right) 
\end{align}
where $\BW^{(m)}$ is a $D \times K$ weight matrix that contains in its columns the contributions to the means for each of the possible configurations of $\Bx_{t}^{(m)}$. The marginal over $\By_t$ is thus a Gaussian mixture model, with $K^M$ Gaussian mixture components, each having a constant covariance matrix $\Sigma$.

The parameters $\boldsymbol{\theta}_{\text{FHMM}} = \{\BP^{1:M}, \BW^{1:M}, \Sigma\}$ can be learned using EM, as before.
In this case, however, the E-step becomes intractable, since the forward-backward step has time complexity $\mathcal{O}(TMK^{M+1})$. An alternative approximation that works well in practice is the structured mean field approximation,
which has time complexity $\mathcal{O}(TMK^2 I)$, where $I$ is the number of mean field iterations \citep[see][for details]{GhahramaniFHMM1997,MurphyBook2012}.

\begin{figure*}[!t]
\begin{center}
  \includegraphics[width=14.0cm]{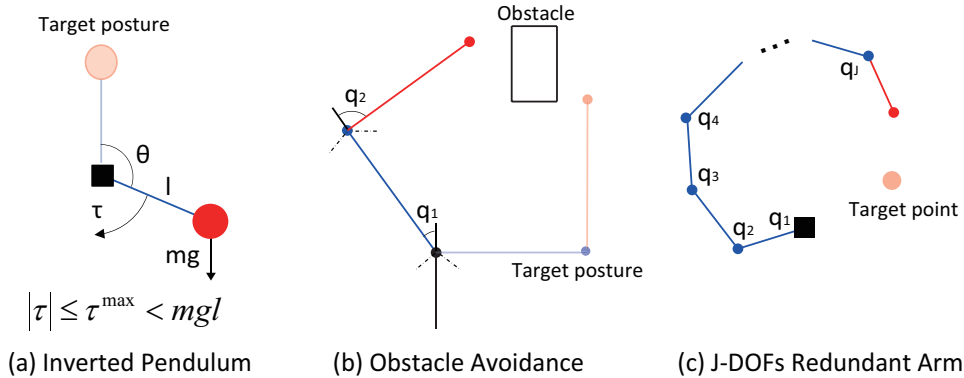}
\caption{
Motor control problems with simulated robots: {\bf (a) Pendulum swing up with limited torque.} 
The state variable is $\ysys = [\theta, \omega]^T$ where $\omega = \dot{\theta}$, $|\theta| \leq \pi$, $|\omega| \leq 4 \pi$. 
The control input is the torque at the joint $\tau$. 
Uncontrolled dynamics and control matrix are given as 
$\Bf(\ysys) = \left[0~1; g \sin(\theta)/l ~ \mu/\omega \right]$, $\BB(\ysys) = 1/ml^2$, respectively. 
Parameters values are $m=l=1$, $g=9.8$, $\mu=0.25$ 
and $\tau^{max} = 5.0$ that satisfies $\tau^{max}< mgl$; 
{\bf (b)~Robot arm control with obstacle.}
The state variable is 
$\ysys = [q_1, q_2]^T \in \mathcal{S}$ where $\mathcal{S}$ is the state space that satisfies the joint angle limits and no collisions 
with the obstacle. The control input is $\Btau = \dot{\ysys}$. 
The uncontrolled dynamics and control matrix are $\Bf(\ysys) = [0,0]^T$ and $\BB(\ysys) = \mathcal{I}_D$; 
{\bf (c) Multi-DOF redundant arm reaching task.} 
The state variable is $\ysys_t = [q_1(t), \hdots,q_J(t)]^T$, $q_i(t) \in \mathcal{S}$ is the $i$-th joint angle and 
$\mathcal{S}$ is the state space that satisfies the joint angle limit $-0.5 \pi \leq q_i(t) \leq 0.5 \pi$.
The control input, uncontrolled dynamics and control matrix are as in (b), but for $J$ dimensions.
In all examples we use first-order Euler method for numerical integration. 
}
\label{Task}
\end{center}
\end{figure*}

\subsection{CONTROL COMPUTATION IN LATENT SPACE}
\label{sec:costfhmm}
In a similar way as in Section \ref{sec:costhmm}, we need first to define a cost function in the latent space $q(\Bx_t)$ to be able to formulate a KL control problem.
A natural way to define $q(\Bx_t)$ given the observation model of Eq.~\eqref{eq:obs-FHMM} and the cost function in observation space $\ycost$ is 
\begin{align}
q(\Bx_t) = \alpha \tilde{q}\left(\sum_{m=1}^M \BW^{(m)} \Bx_t^{(m)}\right).
\end{align}
Computing the exact optimal control using Eq.~(\ref{eq:linear}) in FHMMs requires to transform the model into a single chain model with $K^M$ states, which is intractable. We assume approximate controlled dynamics $u\app(\Bx_{t}|\Bx_{t-1})$ and associated stationary distribution $\Pi\app(\Bx_t)$ that factorize in its components:
\begin{align*}
u\app(\Bx_{t}|\Bx_{t-1}) &= \prod_{m=1}^M u\app^{(m)}(\Bx^{(m)}_{t}|\Bx^{(m)}_{t-1})\notag\\
\Pi\app(\Bx_t) &= \prod_{m=1}^M \Pi\app^{(m)}(\Bx^{(m)}_t). 
\end{align*}
These assumptions imply that the KL cost term can also be decomposed such that Eq.~\eqref{eq:avgcost} becomes
\begin{align}\label{eq:expcostKL}
& \sum_{\Bx_t} \prod_{m=1}^M \Pi\app^{(m)} (\Bx^{(m)}_t)\times \notag\\
& \left( q(\Bx_t) + \sum_{m=1}^M \text{KL}\left( u\app^{(m)}(\cdot|\Bx^{(m)}_t)\left\| p^{(m)}(\cdot|\Bx^{(m)}_t\right.)\right)\right). 
\end{align}

We can minimize Eq.~\eqref{eq:expcostKL} iteratively using sequential updates: for each chain~$m$, update the parameters $u\app^{(m)}$ and $\Pi\app^{(m)}$ assuming the parameters for the other chains fixed so that it minimizes the marginal state-dependent cost 
\begin{align}\label{eq:costVKL}
    Q^{(m)}(\Bx^{(m)}_t)&=\sum_{\Bx_t^{(i)}, i \neq m} \prod_{i \neq m} \Pi^{(i)}\app(\Bx_t^{(i)}) q(\Bx_t)
\end{align} 
and the corresponding KL cost. 
Each update corresponds to a sub-problem of the type of Eq.~\eqref{eq:linear} and can be solved as a principal eigenvector problem. The average cost monotonically decreases at each iteration and its convergence is guaranteed. We call this scheme \emph{Variational KL minimization} {\bf(VKL)}.


Note however, {\bf VKL} requires summing over all the values of the $M-1$ chains to obtain the 
marginal state-dependent cost, and thus it has time complexity $\mathcal{O}(K^{M-1})$, which is still intractable. We further approximate this computation by taking the expected state of the other chains according to their individual stationary distributions
\begin{align}\label{eq:costAVKL}
 Q^{(m)}(\Bx^{(m)}_t) &\approx\alpha\tilde{q}\left( \BW^{(m)}\Bx_t^{(m)} +\sum_{i\neq m} \BW^{(i)}\Pi\app^{(i)} \right),
\end{align}
where $\Pi\app^{(i)}$ is a $K$-dimensional vector with the stationary distribution of chain $i$. Evaluation of Eq.~\eqref{eq:costAVKL} only requires $\mathcal{O}(KM)$ steps, and it is therefore tractable. We refer this approximation as \emph{Approximate Variational KL minimization} {\bf(AVKL)}. 

We refer to the control computed using either {\bf VKL} and {\bf AVKL} as $u\app^ *$ in the rest of this section. 

\subsection{CONTROL COMPUTATION IN OBSERVED SPACE}
\label{sec:obsFHMM}
Having approximated our optimal control law in the latent space, we need to define a control law for the real (observed) system given sequence of observations $\By_{1:t}$. We follow the same approach as in Section~\ref{sec:optcont}. First, we obtain estimates for the expected values of the next observed state under both  controlled and uncontrolled dynamics as $\bar{\By}_{t+1|1:t}^u$ and $\bar{\By}_{t+1|1:t}^p$, respectively. 
Second, we apply the controller of Eq.~\eqref{eq:pd}.

The first step requires to solve a filtering problem to obtain $u(\Bx_t|\By_{1:t})$, which is intractable for this model. We use an approximate approach based on structured mean field, as in the model learning step (Section \ref{subsec41}, E-step). 
However, instead of keeping the last filtered estimate $u(\Bx_{t-1}|\By_{1:t-1})$ as before, we keep the filtered estimate at time-step $t-H$, i.e. $u(\Bx_{t-H}|\By_{1:t-H})$ and perform \emph{offline} structured mean field using the last $H$ observations $\By_{t-H:t}$. This approach improves considerably the accuracy of the filtered estimates $u(\Bx_t|\By_{1:t})=\prod_m u^{(m)}(\Bx_t^{(m)}|\By_{1:t})$ and at the same time, it is more efficient than structured mean field on the entire sequence of past observations.


Once we have filtered estimates of the latent state, the expectation of $\By_{t+1}$ over predictive distribution $u(\By_{t+1}|\By_{1:t})$ can be approximated using samples
\begin{align*}
\bar{\By}^u_{t+1|1:t}=  &\int \By_{t+1} u(\By_{t+1}|\By_{1:t}) d\By_{t+1}\notag\\  \approx & \frac{1}{L} \sum_{\mu=1}^L \sum_{m=1}^M \BW^{(m)} \hat{\Bx}^{(m)}_{\mu}
\end{align*}
where $\hat{\Bx}^{(m)}_{\mu}$ are samples drawn from the posterior distribution of the latent component according to the approximated controlled dynamics
\begin{eqnarray}
\hat{\Bx}^{(m)}_{\mu} &\sim& u^{(m)}(\Bx_{t+1}^{(m)}|\By_{1:t})\nonumber \\ 
&=& \sum_{\Bx_t^{(m)}}u^{*,(m)}_\text{ap}(\Bx_{t+1}^{(m)}|\Bx_t^{(m)}) u^{(m)}(\Bx_t^{(m)}|\By_{1:t}).~~~~
\end{eqnarray}
Similarly, we can estimate $\bar{\By}^p_{t+1|1:t}$ using samples from
\begin{align*}
p^{(m)}(\Bx_{t+1}^{(m)}|\By_{1:t}) & =\sum_{\Bx_t^{(m)}}p^{(m)}(\Bx_{t+1}^{(m)}|\Bx_t^{(m)}) u^{(m)}(\Bx_t^{(m)}|\By_{1:t}). 
\end{align*}


We show in the next section that for relatively small values of the window length $H$ and the number of samples $L$, the resulting controls are satisfactory.

\section{SIMULATION RESULTS}
\label{sec:exp}
In this section, we apply our method to three benchmark (simulated) robot motor control problems:
(a) pendulum swing-up with limited torque \citep{DoyaNC2000}, 
(b) robot arm control with obstacles \citep{SugiyamaICRA2007}, 
and (c) multi-degrees of freedom (DOF) redundant arm reaching task \citep{TheodorouBS10}. Figure \ref{Task} illustrates these problems.
The first two examples correspond to the approach using HMMs of Section \ref{sec:hmm} whereas the third shows an application using FHMMs as described in Section \ref{sec4}.

\begin{figure}[!t]
\begin{center}
  \includegraphics[width=.4\textwidth]{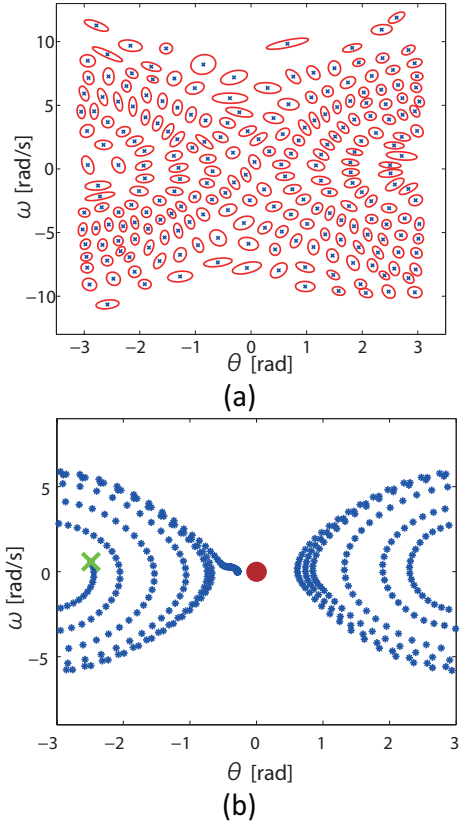}
\caption{Pendulum swing-up task results: {\bf(a)} Observation model after learning the HMM with $N=225$ hidden states and $\sigma_{\epsilon}=1.5$. 
Each hidden state corresponds to a two-dimensional Gaussian distribution with mean indicated by a cross and contour with equal probability density shown as an ellipse.
{\bf (b)} Typical controlled behaviour in the
phase plane. The cross and the circle show initial and target states
respectively.}
\label{Pen2} 	
\end{center}
\end{figure}

For learning the HMM parameters, we use identical and independent exploration noise in all controlled dimensions parameterized by
$\sigma^2_\epsilon$, i.e. $(\Sigma_\epsilon)_{ij}=\delta_{ij}\sigma^2_\epsilon$. Both tasks consider a two-dimensional observed continuous state and a one-dimensional latent variable.  The complexity of the method strongly depends on the number of
hidden values $N$.  For this experiments, we simply choose $N$ large enough ($N=255$ in both scenarios) to obtain a model that accurately describes the system dynamics.  We learn the full parameter vector $\boldsymbol{\theta}_{\text{HMM}}$ using EM with K-means initialization for the Gaussian means.

\begin{figure*}[!t]
\begin{center}
  \includegraphics[width=15.50cm]{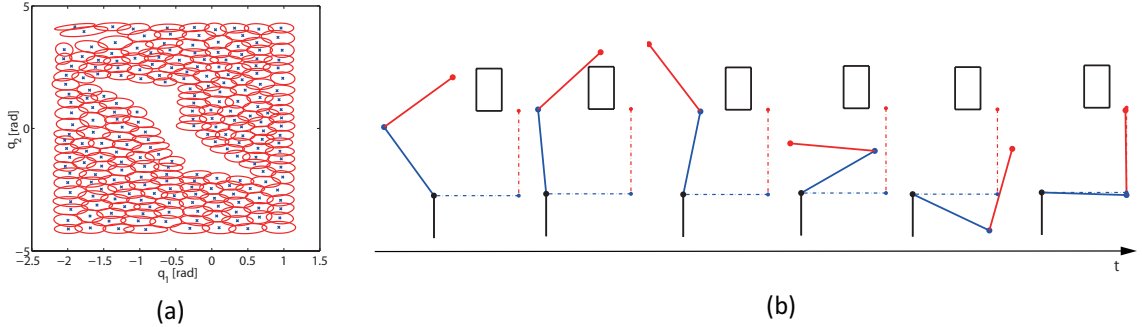}
\caption{Results on the robot arm with an obstacle.  {\bf (a)} Learned HMM
with $N=225$, $\Sigma_{\epsilon} = {\rm diag}\{1.5,
1.5\}$ and $T=3\cdot 10^4$ samples.  {\bf (b)} Controlled robot arm behavior at
different time steps.  The robot successfully reaches the target posture 
avoiding the obstacle. }
\label{Arm}
\end{center}
\end{figure*}
\subsection{PENDULUM SWING-UP TASK}
This is a non-trivial problem when the maximum torque $\tau^{\rm max}$ is
smaller than the maximal load torque $mgl$. The optimal control requires to
take an energy-efficient strategy: swing the pendulum several times to build up
momentum and also decelerate the pendulum  early enough to prevent it from
falling over.

Fig.~\ref{Pen2}(a) shows the 2-dimensional observation model after learning with exploration noise $\sigma^2_{\epsilon}=1.5$. We can see that the HMM is able to capture a discrete, coarse representation of the continuous state.

For control computation, we define a quadratic cost $\ycost= \By ^T {\Sigma_q^{-1}} \By$, where
$\Sigma_q = {\rm diag}\{0.005, 0.02\}$, and set the scale parameter $\alpha=\alpha_0 \Delta t / \sigma^2_{\epsilon}$ to prevent the scaling effect of the
exploration noise variance $\sigma^2_{\epsilon}$ in the KL cost
$(\alpha_0=0.2)$.  The gain matrix is $\BK = {\rm diag}\{50,10\}$.  The
eigenvector computation only takes $3\cdot 10^{-2}$ seconds 
\footnote{Core-i7 2.8GHz-CPU, 8GB memory and MATLAB.}.
The
computation of control input (see Section~\ref{sec:optcont}) takes
$3\cdot 10^{-3}$ seconds per time-step. 
The resulting controller successfully maintains the pendulum in a region of
$|\theta| \leq 0.5$ continuously in all tested random
initializations and it is optimal in terms of energy-efficiency.
A typical controlled behavior of the pendulum is shown in
Fig.~\ref{Pen2}(b). 

For comparison, we also implemented standard value iteration (VI) \citep{SuttonBook1998}, which requires knowledge of the true pendulum dynamics and uses a fully discretized state-action space.  For consistency, we choose as a cost function $r(\By,\Bu) = \alpha \ycost + \frac{1}{2}||\Bu||^2$ and the same error tolerance $10^{-8}$ for both value iteration method and power method.  VI requires a very fine discretization ($N\geq 1225$ states) and at least $20$ seconds of CPU-time, which are roughly an order of magnitude larger than the values obtained using the proposed method. 

\subsection{ROBOT ARM CONTROL WITH OBSTACLE}
In this second task, we aim to control a two-joint robot arm from an initial
posture to the target posture while avoiding an obstacle. The presence of the
obstacle makes this task difficult to solve using standard trajectory
interpolation methods, see Fig.~\ref{Task}(b) for details.

Fig.~\ref{Arm}(a) shows the 2D observation model learned using
the same setup as before. As the empty region in the middle of the plot indicates, the model successfully captures the physically impossible state transitions that would bring the robot arm through the obstacle.

For this problem, we set the cost function as $\ycost= (\By -
\Bg)^T{\Sigma_q^{-1}}(\By - \Bg) $, where $\Sigma_q = {\rm diag}\{0.01, 0.01\}$
and $\Bg = [-\pi/2, \pi/2]^T$.  In this case, we use $\alpha_0=0.05$ and $\BK =
{\rm diag} \{ 3.0,0.5 \}$ to set the scale parameter and the gain matrix,
respectively.  
Computation time of the optimal control is approximately $0.03$ seconds using
the same specifications as in the previous example. 
Fig.~\ref{Arm}(b) illustrates the typical controlled robot behavior.  The robot
arm first decreases the angle $q_2$ and then modifies $q_1$ reaching the target posture while successfully avoiding the obstacle. 

\begin{figure*}[!t]
\begin{tabular}{ccc}
\begin{minipage}{0.32\hsize}
\subfigure[CPU time for KL minimization]{
\centering
\includegraphics[width=1\hsize]{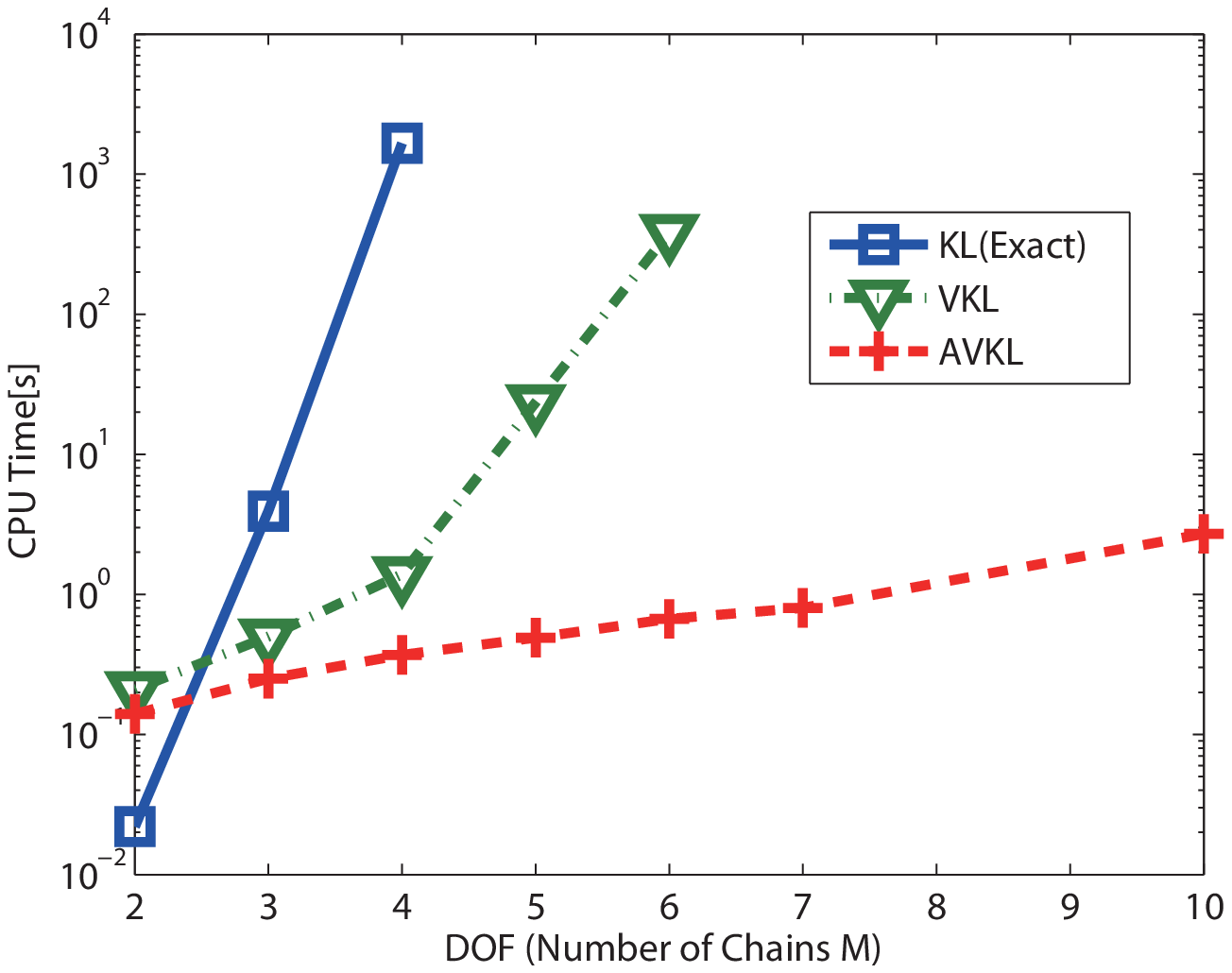}
\label{fig:CPU:1}}
\end{minipage}
\begin{minipage}{0.32\hsize}
\subfigure[Control Error]{
\centering
\includegraphics[width=1\hsize]{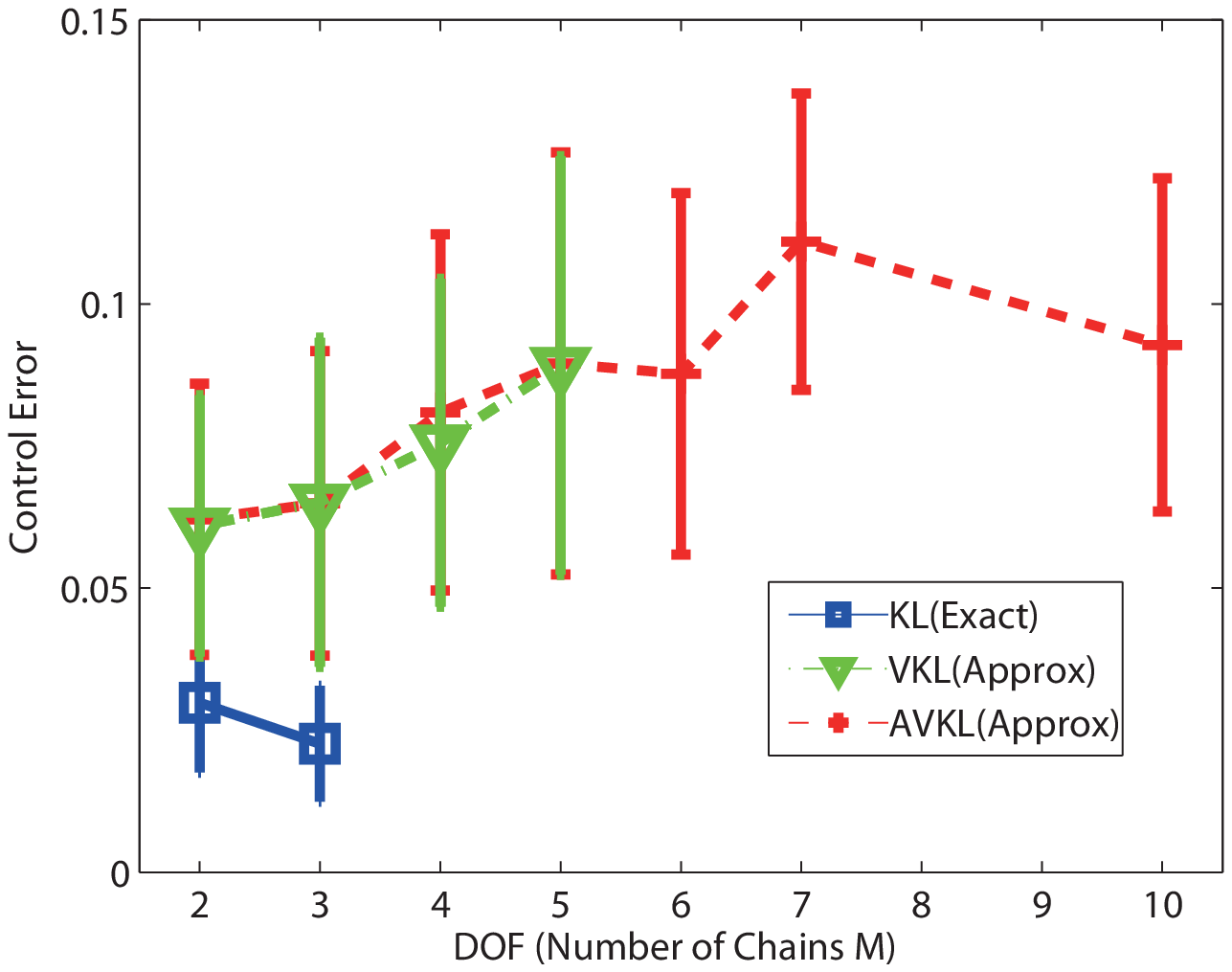}
\label{fig:CPU:2}}
\end{minipage}
\begin{minipage}{0.32\hsize}
\subfigure[CPU time for computing control input]{
\centering
\includegraphics[width=1\hsize]{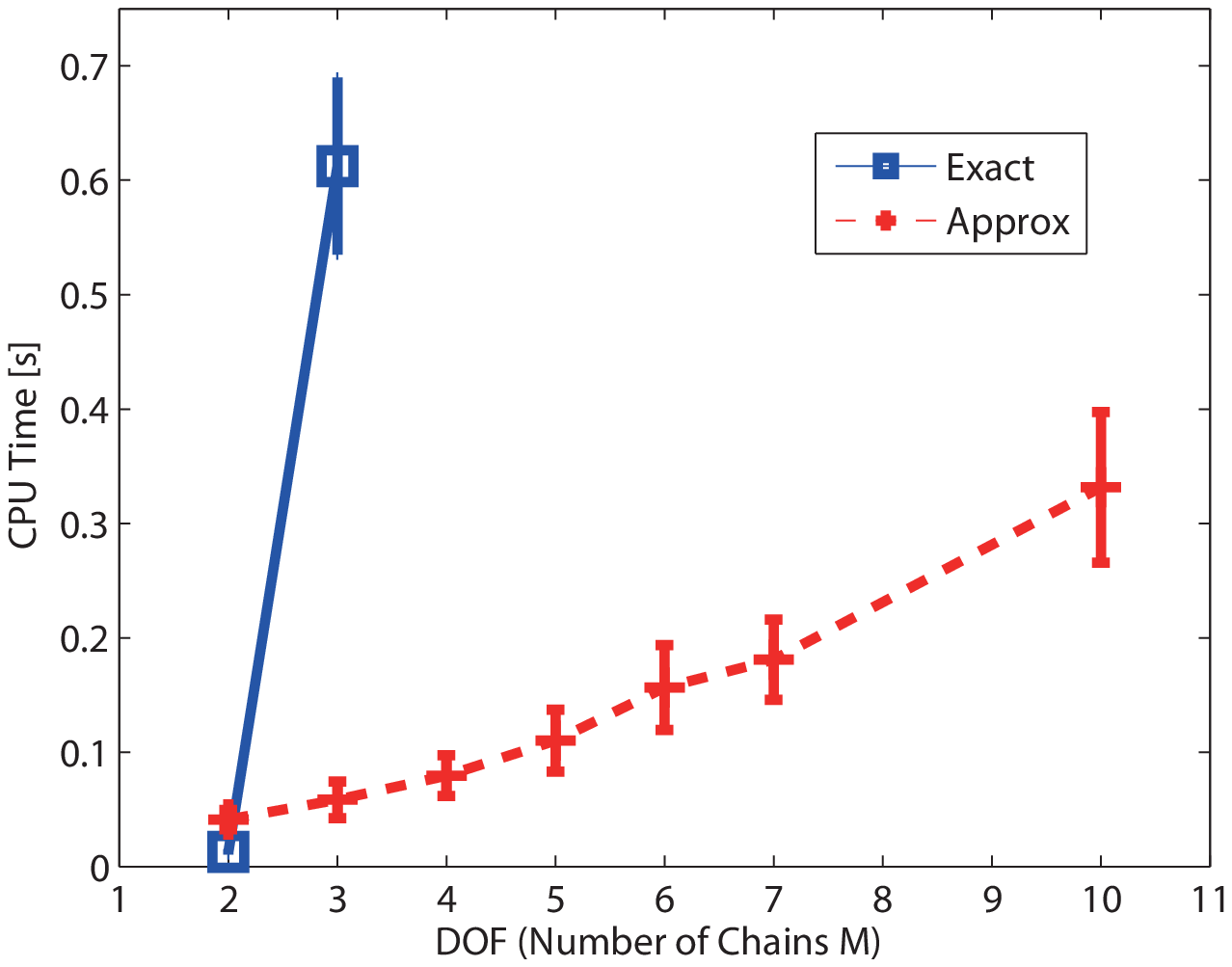}
\label{fig:CPU:3}}
\end{minipage}
\end{tabular}
\caption{
\textbf{Multi-DOF robot reaching task:} Comparison between {\bf KL(exact)}, {\bf VKL} and  {\bf AVKL}. {\bf VKL} and {\bf AVKL} can efficiently compute near optimal controller comparable to exact KL minimization. {\bf AVKL} scales to high-dimensional problems. {\bf KL(exact)} and {\bf VKL} are only feasible for $J<4$ and $J<6$, respectively. }
\label{fig:CPU}
\end{figure*}

\subsection{REACHING TASK}
The third task consists of a multi-DOF planar robot arm with $J$ joints and joint-limit constraints as shown in Fig. \ref{Task}(c). The $J$ joints are of equal length $l=1$ and connected to a fixed base. Each joint dynamics of this robot model is decoupled, and therefore suitable for our method using FHMMs.

The goal is to control the joint angles to reach a target position 
$\Bt^{\rm target}$ with the end-effector of the robot arm. For $J \gg 2$ the control policy has to make a choice among many possible trajectories in the joint space. Moreover, considering joint-limit constraints limits direct application of standard methods for inverse kinematic, e.g. Jacobian inverse techniques \citep{YoshikawaBook}.
The cost function for this task is
\begin{align}\label{eq:errndof}
\tilde{q}(\By) &= \parallel\Bt^{\rm target} - \BT(\By)\parallel,
\end{align}
where $\BT(\cdot)$ is the forward kinematics model that maps a joint angle vector to the corresponding end-effector position in the task space
\begin{align*}
\BT(\By) = \left[ 
    \begin{array}{cc}
      \sum_{n=1}^J \cos\left( \sum_{j=1}^n y_j \right) \\
      \sum_{n=1}^J \sin\left( \sum_{j=1}^n y_j \right)       
    \end{array}
\right].
\end{align*}

Although the dynamics decouples for each joint, the cost function couples all the joint angles making the problem difficult.

We analyze the scaling properties with the number $J$ of degrees of freedom, comparing the different strategies described in Section \ref{sec:costfhmm}: {\bf KL (exact)} minimization, {\bf VKL} and {\bf AVKL}.
The exact solution uses $K^M$ states and performs exact inference.
For approximate methods, we use as many latent dimensions (chains) as joints $M=J$, with $K=20$ and $H=2J$ time-steps for approximate filtering. Note that $M$ could be smaller than $J$, as long as the learned hidden representation captures well the underlying structure and dynamics. We set $M=J$ to simplify the evaluation.

%

Convergence of variational eigen-computations \vkl and \avkl is reached after approximately $10$ iterations in this task (each iteration requires an update of all the parameters of the $J$ joints).
Learning the parameters of the FHMM is sensitive to local minima. In practice, we choose $\BW^{(m)}$ so that each factored state represents each joint dynamics and only learn the uncontrolled dynamics (transition probabilities). Also, $\Bt^{\rm target}$ is set to one of the $\Bw^{(m)}_i$ to prevent space quantization errors in this comparison. 

Fig.~\ref{fig:CPU} illustrates the comparison. Whereas \klexact and \vkl are only feasible for $J<5$ and $J<7$ respectively, \avkl is applicable to a larger number of joints.
Fig.~\ref{fig:CPU}(a) shows CPU-time for control computation in the latent space (Section \ref{sec:costfhmm}), which scales exponentially for both \klexact and \vkl and approximately linear for {\bf AVKL}. 

Fig.~\ref{fig:CPU}(b) shows the error Eq.~\eqref{eq:errndof} averaged over $200$ trials with randomly initialized joints. 
Although exact control computation can be performed for $M<5$, exact inference is only possible for $M<4$.
We can observe that the resulting controls are satisfactory and errors do not differ significantly between \vkl and {\bf AVKL}. Notice that the \avkl error remains approximately constant as a function of $M$.

Fig.~\ref{fig:CPU}(c) shows CPU-time for the control computation in the observed space (Section \ref{sec:obsFHMM}).
While CPU-time for exact computation quickly increases, our approximate approach results in a roughly linear increase. 

\begin{figure*}[!t]
\begin{tabular}{ccc}
\begin{minipage}{0.3\hsize}
\subfigure[2DOF]{
\centering
\includegraphics[width=0.9\hsize]{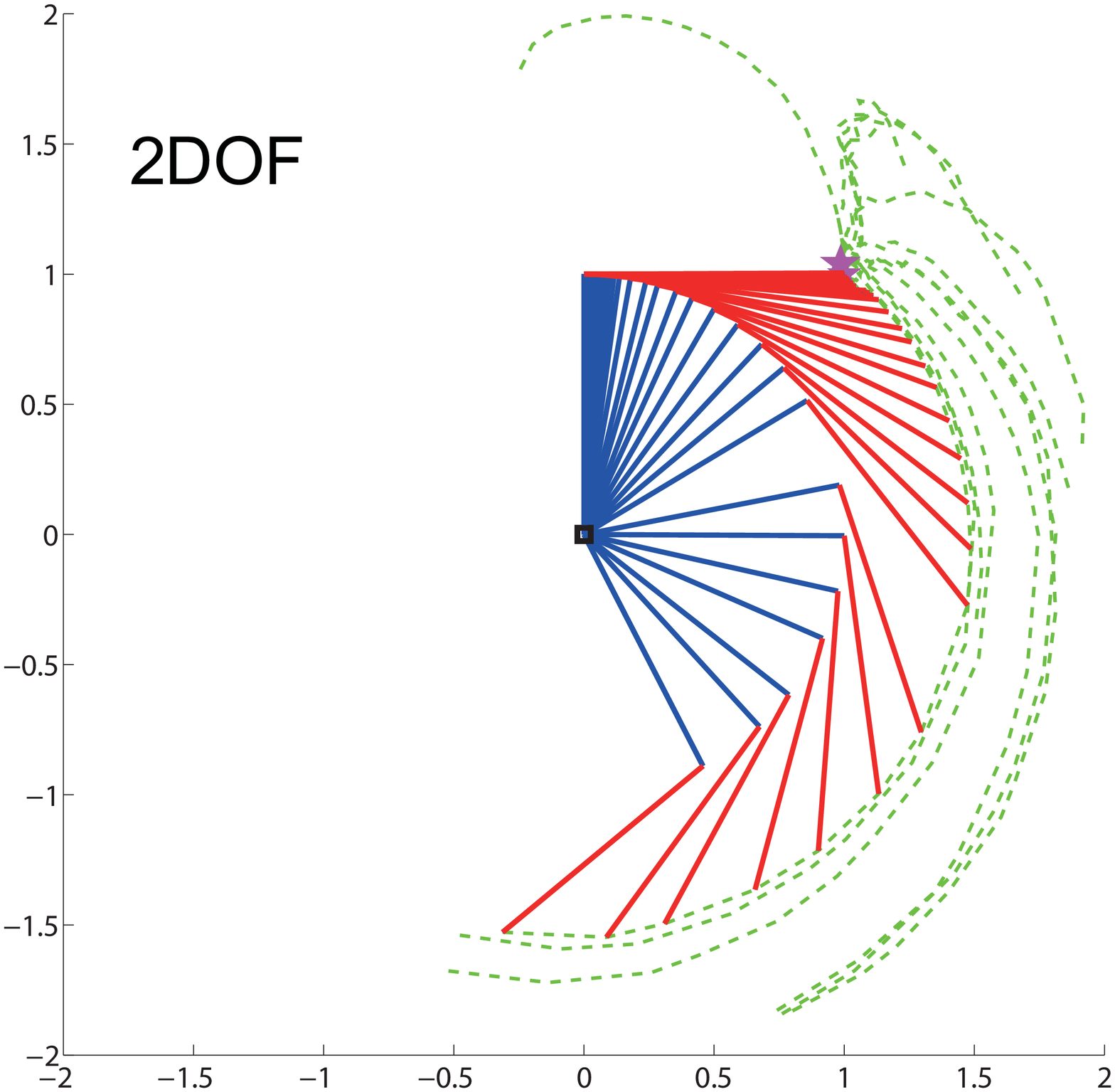}
\label{fig:ARM:1}}
\end{minipage}
\begin{minipage}{0.3\hsize}
\subfigure[10DOF]{
\centering
\includegraphics[width=0.9\hsize]{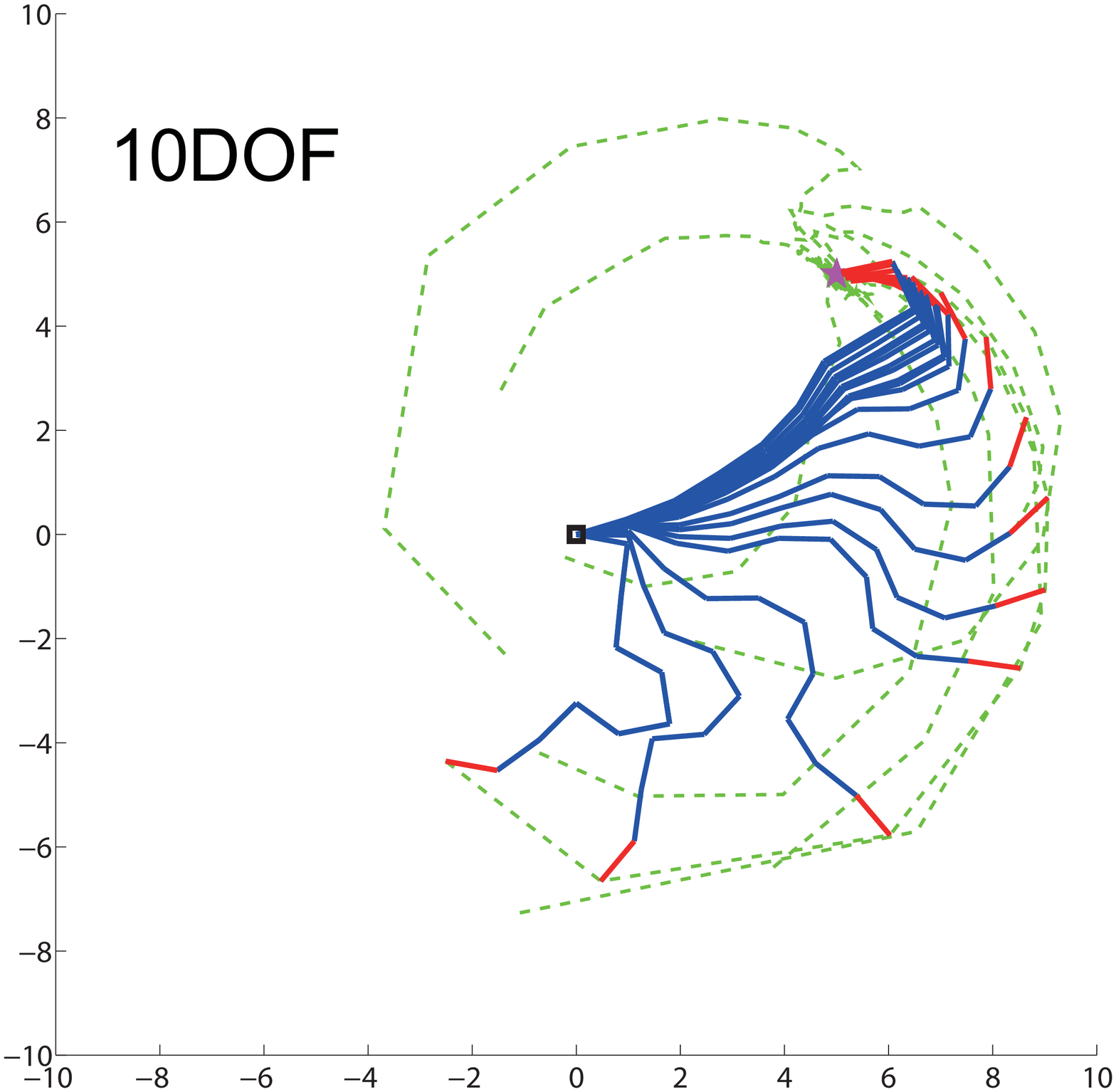}
\label{fig:ARM:2}}
\end{minipage}
\begin{minipage}{0.3\hsize}
\subfigure[25DOF]{
\centering
\includegraphics[width=0.9\hsize]{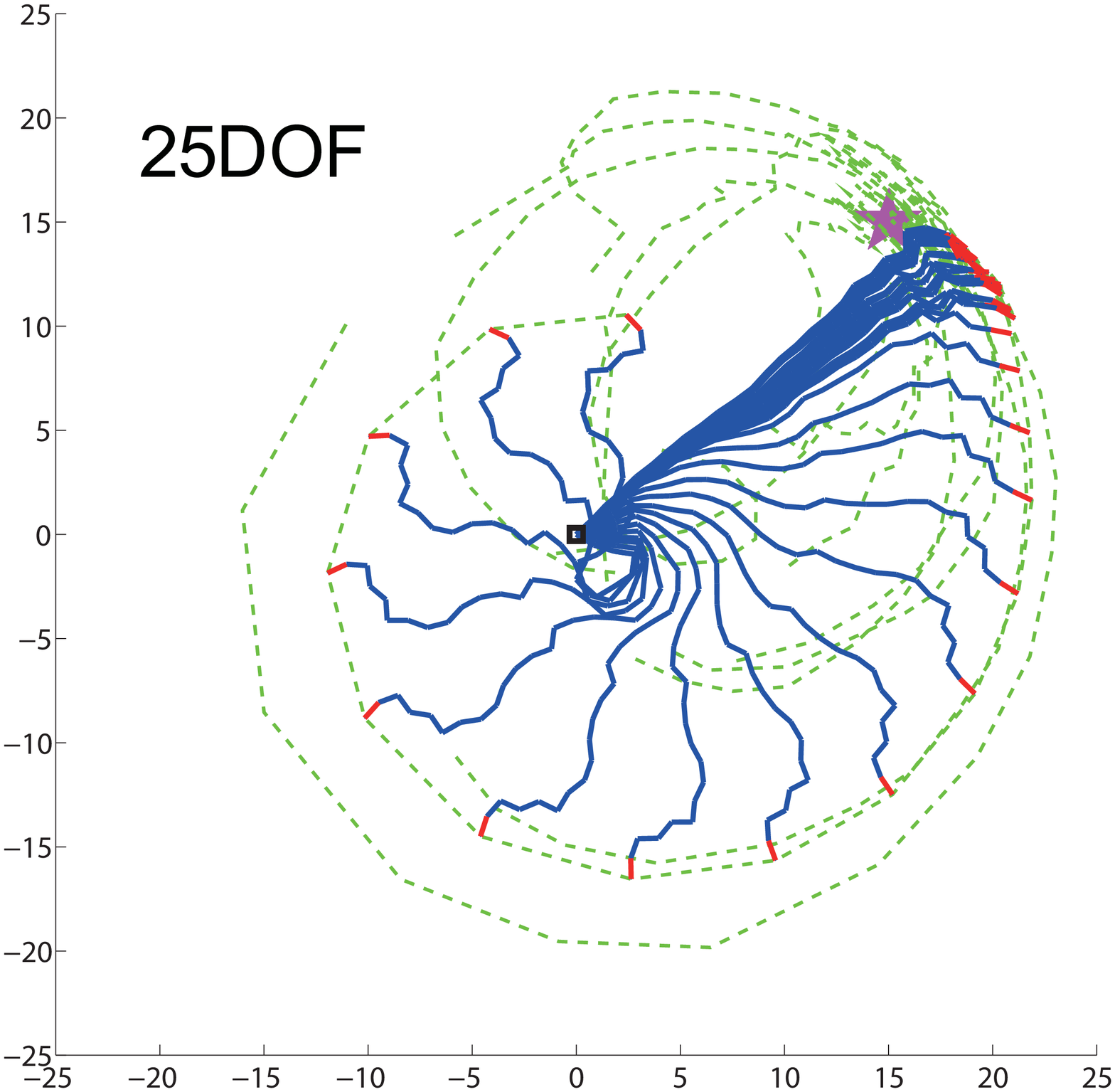}
\label{fig:ARM:3}}
\end{minipage}
\end{tabular}
\caption{
\textbf{Multi-DOF robot reaching task:} Examples of robot trajectories.
The arm successfully reaches the target position while satisfying the joint-limit constraints from several initializations. Green lines show end-effector trajectories for different initializations. Blue and red lines indicate 
intermediate and end links.}
\label{fig:ARM}
\end{figure*}

%
Examples of controlled robot behaviors for a different number of degrees of freedom are shown in Fig.~\ref{fig:ARM}. 
In all cases, the robot successfully reaches the goal while satisfying the joint-limit constraints starting from several initial postures. 

From these results we can conclude that it is feasible to learn FHMMs for high-dimensional systems with uncoupled uncontrolled dynamics and that latent KL control is an effective method to near-optimally control such systems.


\section{DISCUSSION}
\label{sec6}
We have proposed a novel solution that combines the KL control framework with probabilistic graphical models in the infinite horizon, average cost setting. Our approach learns a coarse, discrete representation amenable for efficient computation to near-optimally control continuous-state systems. We have presented two examples, using hidden Markov models (HMMs) and factorial HMMs (FHMMs), and we have shown evidence that our proposed method is feasible in three robotic tasks. In particular, we have demonstrated that the second example with FHMMs is scalable to higher dimensional problems. 

The presented latent KL control approach (with HMMs) resembles the one of \citet{ZhongIFAC2011} which considers an ``aggregated'' space similar to the latent space of the HMM. However, note that 
whereas for \citet{ZhongIFAC2011}  the real model is \emph{required in the observed space},
in our case we \emph{learn} an \emph{approximate} model in which observations are coupled through the latent variables.
Their main computational bottleneck is the ``double'' numerical integration over the observed space for computing the ``aggregated'' state transition probability. In our case, we replace such a problem by a probabilistic graphical model learning problem.

The control performance strongly depends on the quality of the learned model, 
which requires choosing a proper exploration noise and a proper initialization of the graphical model parameters.
Current work is focused in alternative learning methods that efficiently sample interesting regions of the state space and exploit the ergodic nature of the problems. Extension to more complex scenarios is also being considered.

\section*{ACKNOWLEDGEMENT}
This work was supported by JSPS KAKENHI Grant Number 25540085 and by the
European Community Seventh Framework Programme (FP7/2007-2013) under grant
agreement 270327 (CompLACS).


%
%
%
%


{\footnotesize
\bibliography{myrefKL}
 }
\end{document}